\documentclass[a4paper,12pt]{article}
\usepackage{amsmath,amssymb,amsthm}
\usepackage{latexsym,graphicx,color,bbm,subfigure}
\pdfoutput=1
\usepackage{pdfsync}   
\usepackage{bbm}

\newcommand{\be}{\begin{equation}}
\newcommand{\ee}{\end{equation}}
\newcommand{\beq}{\begin{eqnarray}}
\newcommand{\eeq}{\end{eqnarray}}

\newcommand{\bea}{\begin{eqnarray}}
\newcommand{\eea}{\end{eqnarray}}
\def\nin{\noindent}

\begin{document}

\thispagestyle{empty}
\begin{flushright}
IFUP-TH/2012-15
\end{flushright}
\vspace{10mm}
\begin{center}
{\Large \bf Non-Abelian confinement 
and the  dual gauge symmetry:   \\
Many faces of flavor symmetry} 
\\[15mm]
{ Kenichi Konishi$^{a,b}$  
} \footnote{\it e-mail address: 
konishi(at)df.unipi.it   \\
Contribution to QCD12, Montpellier }
\vskip 6 mm

\bigskip\bigskip
{\it  
$^a$  
~Department of Physics ``E. Fermi'', University of Pisa, \\
Largo Pontecorvo, 3, Ed. C, 56127 Pisa, Italy
\\
$^b$
  INFN, Sezione di Pisa,
Largo Pontecorvo, 3, Ed. C, 56127 Pisa, Italy 
  }

\vskip 6 mm

\bigskip
\bigskip

{\bf Abstract}\\[5mm]
{\parbox{14cm}{\hspace{5mm}
\small

We review the physics of confinement based on non-Abelian dual superconductor picture,   
relying on exact solutions in ${\cal N}=2$ supersymmetric QCD and based on the recent developments 
in our understanding of non-Abelian vortices and monopoles. The non-Abelian monopoles, though they are basically just  the 't Hooft-Polyakov $SU(2)$ monopoles embedded in various corners of the larger gauge group, require flavor symmetry  in an essential way for their very existence.  The phenomenon of  flavor-color-flavor separation characterizes  the multiple roles flavor symmetry plays in 
producing quantum-mechanical non-Abelian monopoles.

}}
\end{center}
\newpage
\pagenumbering{arabic}
\setcounter{page}{1}
\setcounter{footnote}{0}
\renewcommand{\thefootnote}{\arabic{footnote}}

\section{Confinement as non-Abelian dual superconductor }
\nin
It has become customary to think of confinement as a dual superconductor \cite{NM,TH}:  dual Higgs mechanism.  It is implicitly assumed in their work that 
the system goes through dynamical Abelianization,
\beq   SU(3) \to U(1)^{2}. \label{DGB}
\eeq
But what is the relation between the (dual) gauge symmetry breaking  (\ref{DGB}) and the chiral symmetry breaking 
\beq    SU_{L}(2) \times SU_{R}(2)  \to  SU_{V}(2)  \label{CSB}
\eeq
?  Lattice simulation indicates that, in theories with flavor in the fundamental representation, the chiral symmetry restoration and deconfinement temperatures coincide:
\beq      T_{c}=T_{\chi}: 
\eeq
this would suggests that the same order parameter
\beq   \langle M \rangle \propto \Lambda  \ne 0,
\eeq
describe both  (\ref{CSB}) and confinement.  But  a condensate of monopoles carrying the  $SU_{L}(2) \times SU_{R}(2) $ quantum numbers \footnote{A monopole can carry  flavor quantum numbers such as in (\ref{suchas}) due to the Jackiw-Rebbi effect \cite{JR}.}
\beq    \langle M_{i}^{j} \rangle \propto    \delta_{i}^{j} \, \Lambda    \label{suchas}
\eeq
where $i,j=1,2$ are the left and right $SU(2)$ indices, would not work:  it would leads to too many Nambu-Goldstone particles, as the effective theory would have an accidental
$SU(N_{f}^{2})= SU(4)$ symmetry. 

\section{Monopoles  in ${\cal N}=2$ theories}
\nin
It is natural to ask how things work in ${\cal N}=2$ supersymmetric theories where many exact results are known on their dynamics \cite{SW1}.
In the $SU(2)$ gauge theories with  ${\cal N}=2$ supersymmetry with $N_{f}$ flavors, light monopoles with nontrivial flavor quantum charges do occur and 
condense,  leading to the (dual) gauge symmetry breaking and dynamical global symmetry breaking, simultaneously. 
In all cases $N_{f}=1,2,3$, results analogous to  (\ref{DGB}) and (\ref{CSB}), curiously, do not lead to any paradox however. 
For example, in $N_{f}=2$ theory, the monopoles are in either 
\beq    ({\underline 2},  {\underline 1})  \quad {\rm or} \quad   ({\underline 1},  {\underline 2})
\eeq
of $SO(4) \sim SU(2) \times SU(2)$; their condensation induces quark confinement and the global symmetry breaking,
$ SU_{1}(2) \times SU_{2}(2) \to SU_{2}(2)$.  It would appear that in $SU(2)$ theories  an accidental symmetry is avoided ``accidentally''!

Massless Abelian monopoles also appear in {\it pure} ${\cal N} =2$  gauge theories, as infrared degrees of freedom.   

In $SU(N)$ gauge theories ($N\ge 3$) with $N_{f}$ flavors the situation is different \cite{APS,CKM}. If the monopoles in the infrared theory were all Abelian, there
{\it would be} problems of too-many-NG-bosons,  as they would be the sole carriers of flavor charges at low-energies. 
Actually the system avoids falling into such an embarrassing situation by generating light {\it non-Abelian} monopoles. For instance, in ${\cal N} =2$, $SU(N)$ gauge theories with 
$N_{f}$ flavors,  a typical confining vacuum is characterized by the set of non-Abelian and Abelian monopoles listed in Table~\ref{tabnonb} taken from \cite{APS}.  
\begin{table}[b]
\begin{center}
\vskip .3cm
\begin{tabular}{|cccccc|}
\hline
&   $SU(r)  $     &     $U(1)_0$    &      $ U(1)_1$
&     $\ldots $      &   $U(1)_{N-r-1}$      \\
\hline
$n_f \times  {\cal M}$     &    ${\underline {\bf r}} $    &     $1$
&     $0$
&      $\ldots$      &     $0$                \\ \hline
$M_1$                 & ${\underline {\bf 1} } $       &    0
&
1      & \ldots             &  $0$                 \\ \hline
$\vdots $  &    $\vdots   $         &   $\vdots   $        &    $\vdots   $
&             $\ddots $     &     $\vdots   $       
\\ \hline
$M_{N-r-1} $    &  ${\underline {\bf 1}} $    & 0                     & 0
&      $ \ldots  $            & 1                 \\ \hline
\end{tabular}
\caption{The massless non-Abelian and Abelian monopoles  and their charges  at the $r$ vacua
at the root of a ``non-baryonic'' $r$-th   Higgs branch.  }
\label{tabnonb}
\end{center}
\label{Qnumbers}
\end{table}

First of all, we note that ``colored dyons''  do exist (see below), and appear as infrared degrees of freedom. Furthermore these non-Abelian monopoles ${\cal M}$ carry flavor quantum numbers of $SU(N_{f})$, ${\underline N_{f}}$.   Finally, their  vacuum expectation value (VEV) breaks the global symmetry as 
\beq   SU(N_{f})\times U(1) \to   U(r) \times U(N_{f}-r),  
\eeq
as the system in the $r$ vacua is in (dual) color-flavor locked phase  ($ i=1,2, \ldots,  N_{f}, \quad  \alpha=1,2,\ldots, r$), 
\beq     \langle {\cal M}^{i}_{\alpha} \rangle =  v \, \delta^{i}_{\alpha}\;.
\eeq
Our main concern is to understand the quantum numbers carried by the non-Abelian monopoles ${\cal M}$. 

But ``non-Abelian monopoles'' \cite{GNO}-\cite{EW} have notorious problems \cite{CDyons}-\cite{EWYi}!  
Basically they are just the 't Hooft-Polyakov monopoles of gauge symmetry breaking $SU(2)\to U(1)$,  embedded 
in a larger gauge symmetry breaking system, 
\beq  G \to H\;, 
\eeq
for instance  $G= SU(N+1)$ color group, broken by some set of adjoint scalar field to $H=  SU(N) \times U(1)$. 
In such a situation it is natural to think the set of degenerate monopoles lying in various corners of $G$ color group, 
related by the unbroken $H= SU(N)$ as something forming a multiplet of the latter.   This straightforward idea however faces the well-known ``no-go'' theorems:  
\begin{description}
  \item[(i)] Topological obstructions (i.e., in the presence of a monopole the unbroken  $H$  (e.g., $SU(N)$) cannot be defined globally in all directions) \cite{CDyons}. 
    \item[(ii)]  There are certain gauge zeromodes which are non-normalizable \cite{DFHK}: this can be regarded as an infinitesimal version of the problem (i); 
  \item[(iii)]  One might try to approach the problem starting from the totally Abelian breaking $ SU(N+1) \to   U(1)^{N}\;$  by tuning the adjoint VEVs
  so as to reach degenerate diagonal elements.  This approach fails as the procedure necessarily produces ``colored clouds'',  infinitely large and infinitely thin 
  ``monopole'' configurations \cite{EWYi} which defy any consistent treatment. 
   \end{description}
   
\nin But the issue of the foremost importance is the fact that  GNO quantization condition \cite{GNO}-\cite{EW} \footnote{$\alpha$'s are the root vectors of the ``unbroken'' gauge group, 
$\beta$ are the constant vectors characterizing the solutions.}
\beq    2 \, \beta \cdot \alpha \in   {\mathbf Z}   
\eeq 
   implies that the monopoles are to transform according to (if any) representations of the {\it dual} of the ``unbroken'' gauge group $H$,  ${\tilde H}$,   not under $H$ itself. As the GNO  duality is of electromagnetic type, the monopoles are to transform under some field transformations nonlocal with respect to  the original $H$ ransformations. 

Finally there is the question of the phase \cite{Duality}: in order to see the dual group  ${\tilde H}$ in action (unbroken),  e.g., in confinement phase, the system must be studied in a 
Higgs phase of $H$, that is, we are to study the whole system such that at lower energies the group $H$ is completely broken (Higgs phase).  In the example mentioned above,
the gauge symmetry breaking sequence is 
 \beq   SU(N+1) \to  SU(N)\times U(1) \to  {\mathbf 1}\;.     \label{Hierarchy}
 \eeq

So how do the ${\cal N}=2$ gauge theories manage to generate quantum mechanical non-Abelian monopoles? 

\section{Hierarchical gauge symmetry breaking and color-flavor locking} 

To see how they do it is  best to study the hierarchical gauge symmetry breaking, (\ref{Hierarchy}). 
It is indeed sufficient to analyze the same system of softly broken ${\cal N}=2$ QCD, at large adjoint mass $\mu$ and quark masses $m$ \cite{ABEK}. In particular, by choosing 
\beq   m  \gg \mu \gg \Lambda, 
\eeq
the scalar VEV's are then
\beq  \langle \Phi \rangle  =   v_{1} \, \left(\begin{array}{cc}{\bf 1}_N & 0 \\0 & -1 \end{array}\right)\;,  \label{PhiVEV}
\eeq
\beq   \langle Q \rangle  =   v_{2} \, {\bf 1}_{N} \label{SquarkVEV}
\eeq
where $v_{1}= -m$, $v_{2}= \sqrt{\mu m}$,  $v_{1}\gg v_{2}$, and the squark fields $Q$  are written in color (vertical)-flavor (horizontal) mixed matrix form. 
A monopole lies in one of the $SU(2)$ groups broken to $U(1)$,  as the result of the symmetry breaking $SU(N+1)\to  SU(N)\times U(1)$ (see  (\ref{PhiVEV})). 
For instance   $A_{i}$ and $\Phi$  fields can be taken to be the 't Hooft-Polyakov monopole living in the  $(1, N+1)$ corners.

On the other hand the smaller squark VEV breaks the gauge symmetry completely, at the same time maintaining the flavor-color diagonal $SU(N)$ symmetry intact.  
As $\Pi_{1}(SU(N)\times U(1))={\mathbf Z}$  vortices are generated. 

As the underlying system $SU(N+1)$ is simply connected, such a vortex must end: the endpoints are precisely the monopoles (which,  in turn, cannot be stable as $\Pi_{2}(SU(N+1))={\mathbf 1}$).
As the individual vortex breaks the global color-flavor locked $SU(N)$ symmetry to $SU(N-1)\times U(1)$, it possesses orientational zeromodes, generating the motion in the coset, 
\beq     SU(N)/SU(N-1)\times U(1) = CP^{N-1}\;. 
\eeq
These are non-Abelian vortices, found in 2003 \cite{HT,ABEKY} and studied extensively since then \cite{SY}-\cite{GJK}.

\section{A new exact local symmetry for the monopole} 

\begin{figure}
\begin{center}
\includegraphics[width=4.7in]{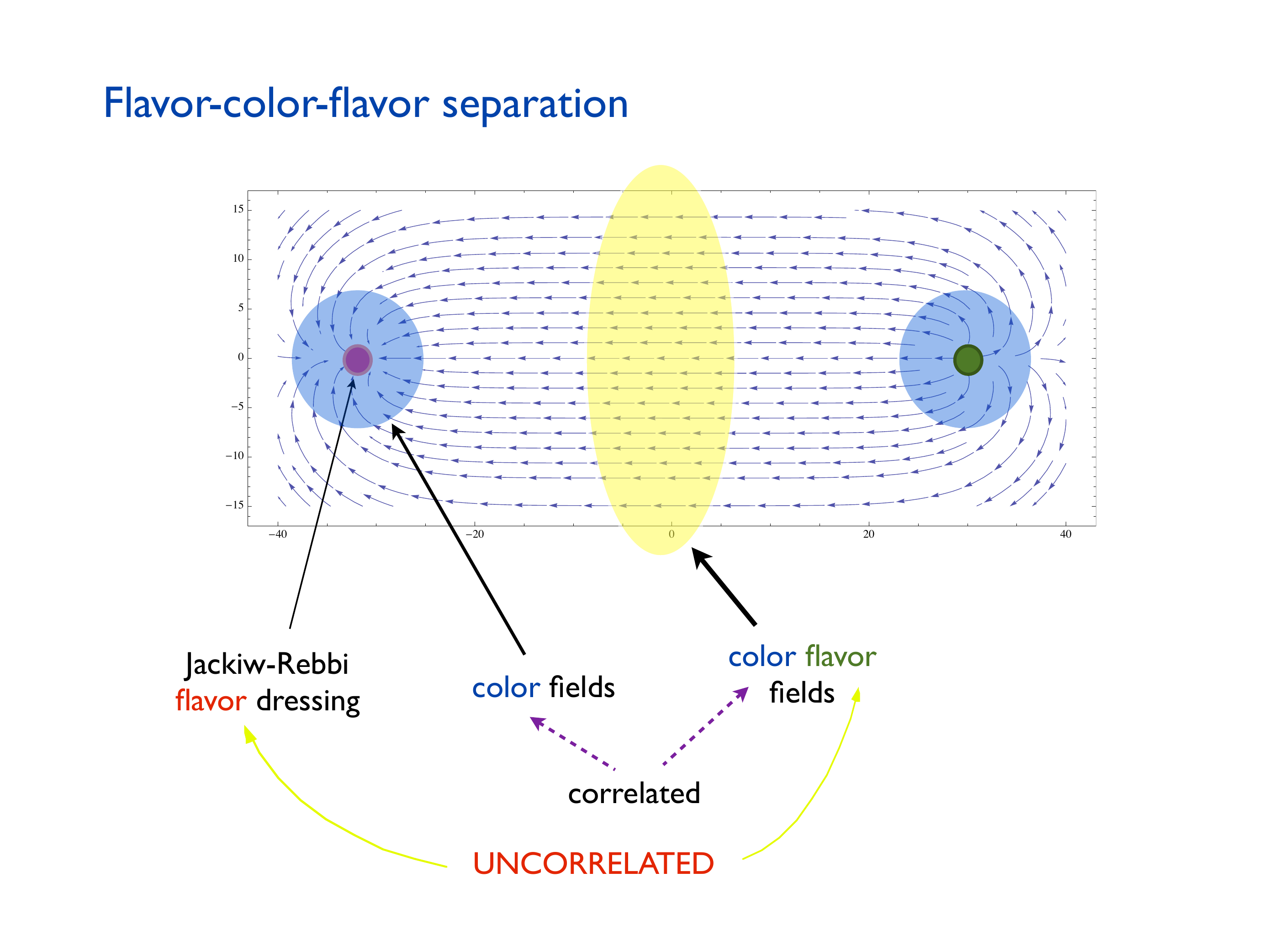}
\caption{ }
\label{FCFsepar}
\end{center}
\end{figure}

\nin To minimize the energy of the whole monopole-vortex complex (see Fig.~\ref{FCFsepar})  the orientation of the monopole (in the color $SU(N)$ space) and of the vortex 
(in the color-flavor diagonal $SU(N)$ space)  must be locked \cite{Cipriani}. As the vortex is oriented in a direction in the color-flavor space (for instance, $(1,1)$ direction), the original $H$ group is {\it explicitly} broken.  Attempt to rotate the monopole in the color $SU(N+1)$ space (the naive non-Abelian monopole concept) would distort  the monopole-vortex complex and raise its energy:  it is no longer a zero-energy mode.
Vice versa,  the whole complex maintains its energy if the monopole (in color)  and  vortex (in color-flavor) are rotated simultaneously. 
 
 But that means that the orientational zeromodes of $CP^{N-1}$ are attached also to the monopoles sitting at the extremes of the vortex.  The monopoles acquire continuous $CP^{N-1}$ zeromodes: they transform as in the fundamental representation of the {\it isometry  group}  of 
$CP^{N-1}$: an $SU(N)$ group.    Note that this $SU(N)$ group is not the original $H$: it is induced  by the original $SU(N+1)\times SU(N)$  color flavor transformations of the original fields, but refers to  motion in the solution space $CP^{N-1}$, which happens to have this form  \footnote{It appears that the fact that the isometry group of the coset $G/H$ is the same as $H$ as a group, in the case of $SU$ theories 
helped obscuring the problem of non-Abelian monopoles in the past.   In an analogous study starting from $G= USp(2N)$ or $SO(2N)$, etc.,  the isometry groups of the cosets, 
of the color-flavor symmetry breaking,   $USp(2N)/U(N)$ or $SO(2N)/U(N)$,  are different  from $H$  as groups  (spinor orbits of $SO(2N)$ or $USp(2N)$, respectivey)}
. 

 We identify this new  group as the dual gauge group ${\tilde H}$. 
  
   We may say that the non-normalizable $3D$  gauge zeromodes of the monopole has been converted  into normalizable $2D$ zeromodes, allowing the monopole to fluctuate in the dual group ${\tilde H}$, but in a confined mode  (the flux does not spread all over $R^{3}$ but can fluctuate and propagate along the vortex: the confining string!).

\section{Many faces of flavor symmetry}
 
\nin   The flavor symmetry thus plays crucial role in the generation of the dual gauge symmetry.  The ``forbidden'' color modulation of the monopole is smoothly converted to the fluctuation of color-flavor orientational modes, which can propagate along the vortex to which the monopole is attached.  We interpret these as the fluctuation of the dual local gauge symmetry, in confinement phase.

Each monopole carries furthermore flavor quantum number of the original $SU(N_{f})$, thanks to the Jackiw-Rebbi mechanism \cite{JR}: dressing by the quark (fermion) zeromodes. These modes are $3D$ normalizable, and do not propagate. The JR effect gives rise to a global charge, in contrast to the 
color-flavor orientational modes.  

The flavor multiplicity of monopoles are essential for the latter to be able to appear as the infrared degrees of freedom, 
in the quantum regime $m, \mu \ll \Lambda $.  

Thus the monopole acquires the dual gauge charge {\it and}  global flavor charge, both related to the original flavor symmetry,  in association with the  gauge dynamics of the soliton monopoles and vortices. The monopole flavor quantum number is  unrelated  to and independent of the dual gauge charge (see Table~\ref{Qnumbers}).  

We call this phenomenon  {\it  flavor-color-flavor separation}. 
 
The main property responsible for this  is the vanishing of the squark fields at the monopole center.  Naturally they vanish all along the vortex core, and this turns out to be essential \cite{Cipriani} for allowing the monopole configuration to be smoothly connected to the vortex.  

\section{Real-world QCD}
\nin
So what is the lesson for the real world QCD? If the chiral symmetry breaking and confinement are caused by the same set of condensates, as suggested by
lattice simulations, there could be two possible dynamical scenarios:
 \begin{description}
  \item[(i)]   Abelian monopoles of dynamically Abelianized $SU(3) \to U(1)^{2}$ gauge group, carrying the $SU_{L}(2) \times SU_{R}(2)$  quantum numbers, condense
  \beq     \langle M^{i}_{j} \rangle = \delta^{i}_{j} \, \Lambda \ne 0;   
  \eeq
  or
  \item[(ii)]  Non-Abelian monopoles $M$, ${\tilde M}$ associated with the dynamical gauge symmetry breaking $SU(3) \to U(2)$, and carrying respectively $SU_{L}(2)$ or $SU_{R}(2)$ flavor quantum numbers,  form composite and condense 
   \beq     \langle M^{i}   {\tilde M}_{j} \rangle = \delta^{i}_{j} \, \Lambda^{2}  \ne 0\;.    
  \eeq
\end{description}   
Scenario (ii) appears to be preferred from the point of view of the correct flavor symmetry realization, avoiding the production of the too-many-NG-bosons in scenario (i).

If we define the homotopy index  $D$  by
\beq   \Pi_{1}(G_{eff}) = {\mathbf Z}^{D}\, 
\eeq 
where $G_{eff}$ is the low-energy effective gauge group, $D=2$ and $D=1$ for the possibilities (i) and (ii), respectively. 
Can one measure $D$ with lattice simulations?

\section*{Acknowledgements}
\nin
We thank the organizers of QCD12, Montpellier, where this talk was presented.


\begin{thebibliography}{999}
\vspace*{-0.25cm}

\bibitem{NM} Y. Nambu, {Phys. Rev.} {\bf  D10}  (1974) 4262,
   S. Mandelstam, {Phys. Lett.} {\bf 53B}  (1975) 476;
                               {Phys. Rep.} {\bf 23C} (1976) 245.
                               
\bibitem {TH}  G. 't Hooft, {Nucl. Phys.} {\bf  B190}   (1981) 455.

\bibitem{JR}
R. Jackiw and C. Rebbi, Phys. Rev. {\bf D13} (1976) 3398.

\bibitem{SW1}
N. Seiberg and E. Witten, Nucl.Phys. {\bf B426} (1994) 19; Erratum
\textit{ibid.} \textbf{B430} (1994) 485, hep-th/9407087,    Nucl. Phys. {\bf B431} (1994) 484,
   hep-th/9408099.

\bibitem{SUN}
 P.  C.  Argyres, M.  R.  Plesser and A.  D.  Shapere, Phys.  Rev.
Lett.  {\bf 75} (1995) 1699, hep-th/9505100; P. C. Argyres and A. D. Shapere, Nucl. Phys. {\bf B461} (1996) 437,
hep-th/9509175;   A. Hanany, Nucl.Phys. {\bf B466} (1996) 85,  hep-th/9509176.


\bibitem{APS}
P.~C.~Argyres, M.~Plesser and N.~Seiberg,
  Nucl.\ Phys.\  {\bf B471} (1996) 159
  [arXiv:hep-th/9603042]. 
  
  \bibitem{CKM}
 G.~Carlino, K.~Konishi and H.~Murayama,
  Nucl.\ Phys.\  {\bf B590} (2000) 37
  [arXiv:hep-th/0005076]. 


\bibitem{GNO} P. Goddard, J. Nuyts and D. Olive,  Nucl. Phys.  B125
(1977) 1.

\bibitem{BS}  F.A. Bais, { Phys. Rev. D18}  (1978) 1206.

\bibitem{EW}   E. J. Weinberg, { Nucl. Phys. B167} (1980) 500;  { Nucl. Phys. B203}  (1982) 445.


\bibitem{CDyons} A. Abouelsaood, { Nucl. Phys. B226} (1983) 309; P. Nelson and A. Manohar,  { Phys. Rev. Lett. 50}
(1983) 943;  A. Balachandran, et. al.,  { Phys. Rev. Lett. 50}
(1983) 1553. 


\bibitem{DFHK} N.~Dorey, C.~Fraser, T.J.~Hollowood and M.A.C.~Kneipp,
  { Phys.Lett. B383}  (1996) 422 [arXiv:hep-th/9605069].

\bibitem{EWYi}  K. Lee, E. J. Weinberg and P. Yi,  { Phys. Rev. D 54 } (1996) 6351  [arXiv:hep-th/9605229].

\bibitem{Duality}
 M.~Eto, et. al.,  
Nucl. Phys. B  {\bf 780} 161-187, 2007
   [arXiv:hep-th/0611313].


\bibitem{ABEK}
  R.~Auzzi, S.~Bolognesi, J.~Evslin and K.~Konishi,
  Nucl.\ Phys.\  B {\bf 686} (2004) 119
  [arXiv:hep-th/0312233];  

  \bibitem{HT}
A.~Hanany and D.~Tong,
JHEP {\bf 0307}, 037 (2003)
[arXiv:hep-th/0306150].


\bibitem{ABEKY}
R.~Auzzi, S.~Bolognesi, J.~Evslin, K.~Konishi and A.~Yung,
Nucl.\ Phys.\ B {\bf 673} (2003) 187
[arXiv:hep-th/0307287].
 
\bibitem{SY}
  M.~Shifman and A.~Yung,  Phys. Rev.  D {\bf 70},  045004  (2004)  [arXiv:hep-th/0403149];  
  
\bibitem{ModMat}
 M.~Eto, Y.~Isozumi, M.~Nitta, K.~Ohashi and N.~Sakai,
 J.\ Phys.\ A  {\bf 39}, R315 (2006)
 [arXiv:hep-th/0602170]. 

\bibitem{GJK} S.B.~Gudnason, Y.~Jiang and  K.~Konishi,   JHEP {\bf 1008}:012 (2010) 
[arXiv:1007.2116 [hep-th]].

\bibitem{Cipriani} 
  M.~Cipriani, D.~Dorigoni, S.~B.~Gudnason, K.~Konishi and A.~Michelini,
  Phys.\ Rev.\ D {\bf 84}, 045024 (2011)
  [arXiv:1106.4214 [hep-th]].



\end{thebibliography}
\end{document}